\documentstyle[11pt]{article}

\makeatletter
\@addtoreset{equation}{section}
\makeatother

\def\dalemb#1#2{{\vbox{\hrule height .#2pt
        \hbox{\vrule width.#2pt height#1pt \kern#1pt
                \vrule width.#2pt}
        \hrule height.#2pt}}}

\let\a=\alpha \let\b=\beta   \let\e=\epsilon
  \let\q=\theta

\def\nn{\nonumber} \def\bd{\begin{document}} \def\ed{\end{document}}
\def\ds{\documentstyle} \let\fr=\frac \let\bl=\bigl \let\br=\bigr
\let\Br=\Bigr \let\Bl=\Bigl 
\let\bm=\bibitem
\let\na=\nabla
\let\pa=\partial \let\ov=\overline
\def\ie{{\it i.e.\ }} 
\newcommand{\be}{\begin{equation}} 
\newcommand{\ee}{\end{equation}} 
\def\ba{\begin{array}}
\def\ea{\end{array}}
\def\ft#1#2{{\textstyle{{\scriptstyle #1}\over {\scriptstyle #2}}}}
\def\fft#1#2{{#1 \over #2}}
\def\del{\partial}
\def\sst#1{{\scriptscriptstyle #1}}
\def\oneone{\rlap 1\mkern4mu{\rm l}}
\def\e7{E_{7(+7)}}
\def\td{\tilde}
\def\wtd{\widetilde}
\def\im{{\rm i}}
\def\bog{Bogomol'nyi\ }
\def\q{{\tilde q}}
\newcommand{\ho}[1]{$\, ^{#1}$}
\newcommand{\hoch}[1]{$\, ^{#1}$}
\newcommand{\bea}{\begin{eqnarray}} 
\newcommand{\eea}{\end{eqnarray}} 
\newcommand{\ra}{\rightarrow}
\newcommand{\lra}{\longrightarrow}
\newcommand{\Lra}{\Leftrightarrow}
\newcommand{\ap}{\alpha^\prime}
\newcommand{\bp}{\tilde \beta^\prime}
\newcommand{\tr}{{\rm tr} }
\newcommand{\Tr}{{\rm Tr} } 
\newcommand{\NP}{Nucl. Phys. }
\newcommand{\tamphys}{\it Center for Theoretical Physics,
Texas A\&M University, College Station, Texas 77843}
\newcommand{\ens}{\it Laboratoire de Physique Th\'eorique de l'\'Ecole
Normale Sup\'erieure\hoch{3}\\
24 Rue Lhomond - 75231 Paris CEDEX 05}

\newcommand{\auth}{H. L\"u\hoch{\dagger1}, S. Mukherji\hoch{\ddagger2} and 
C.N. Pope\hoch{\ddagger1}}

\begin{document}
\begin{flushright}
\hfill{CTP TAMU-65/96}\\
\hfill{LPTENS-96/73}\\
\hfill{hep-th/9612224}\\
\hfill{December 1996}\\
\end{flushright}

\vspace{20pt}

\begin{center}
{\large {\bf From $p$-branes to Cosmology}} 

\vspace{30pt}

\auth

\vspace{15pt}
{\hoch{\dagger}\ens}

\vspace{10pt}
{\hoch{\ddagger}\tamphys}

\vspace{40pt}

\underline{ABSTRACT}
\end{center}

     We study the relationship between static $p$-brane solitons and
cosmological solutions of string theory or M-theory.  We discuss two
different ways in which extremal $p$-branes can be generalised to
non-extremal ones, and show how wide classes of recently discussed
cosmological models can be mapped into non-extremal $p$-brane
solutions of one of these two kinds.  We also extend previous
discussions of cosmological solutions to include some that make use of
cosmological-type terms in the effective action that can arise from the
generalised dimensional reduction of string theory or M-theory.

{\vfill\leftline{}\vfill
\vskip	10pt
\footnoterule
{\footnotesize
	\hoch{1}	Research supported in part by DOE 
grant DE-FG05-91-ER40633 \vskip	-12pt}  \vskip	10pt
{\footnotesize 
        \hoch{2} Research supported in part by NSF grant PHY-9411543 
               \vskip -12pt}  \vskip   10pt

{\footnotesize
        \hoch{3} Unit\'e Propre du Centre National de la Recherche
Scientifique, associ\'ee \`a l'\'Ecole Normale Sup\'erieure et \`a
l'Universit\'e de Paris-Sud}
}

\pagebreak
\setcounter{page}{1}

\section{Introduction}

     A considerable literature exists on the subject of $p$-brane solutions
of the low-energy limits of string theories and M-theory.  These
include the BPS-saturated classes of extremal $p$-branes, which
preserve some fraction of the spacetime supersymmetry, and also their
non-extremal generalisations, where supersymmetry is completely
broken.  A characteristic feature of all these solitonic solutions is
that they are static (or at least stationary, if one allows rotating
solutions as well).  Other studies have focussed on a different class
of solutions to the low-energy effective equations of motion, namely
cosmological solutions in which the metric evolves in time.
Typically, these solutions describe the evolution of a universe from a
small initial radius to a large radius at late times.

     Although they are ostensibly very different in their interpretation,
it has been observed that there are actually many similarities between
some of the $p$-brane solutions and the cosmological solutions
\cite{low1}.  Indeed in some examples, an exact correspondence has
been established \cite{lw}.  The most
obvious difference between the two classes of solution is that the
fields of the cosmological solutions depend upon time, while in the
$p$-brane solitons they depend instead on the coordinates of the
transverse space (in fact, in the cases of relevance here, they depend
just on the radial distance from the origin in the transverse space).
However, this distinction is not such a profound one; it is, for
example, well known that in the interior region of the Schwarzschild
black hole the usual time coordinate becomes spacelike, while the
usual radial coordinate acquires a new interpretation as the time
coordinate \cite{os}.  Thus one way, at least, in which an equivalence
can arise is that the interior solution of a black $p$-brane may be
re-interpreted as a cosmological model.  As we shall see later, not
all of the mappings between $p$-brane and cosmological solutions are
of this kind, and sometimes it is necessary to make Wick rotations in
order to interchange the timelike and spacelike character of
coordinates in the two cases.

     The purpose of this paper is to investigate the mapping between
$p$-brane solutions and cosmological solutions in a somewhat general
framework.  In order to do so, we shall begin by first obtaining a
large class of cosmological solutions. In general, our starting point
can be taken to be a theory describing the coupling of gravity to a
dilatonic scalar field and an antisymmetric tensor field strength of
degree $n$.  A special case of this, which has not been extensively
discussed in the literature, is when the degree $n$ is actually zero.
In this case the ``field strength'' is really just a constant; if
there were no coupling to a dilaton, it would in fact correspond
precisely to a pure cosmological constant term.  In section 2, we
begin by setting up appropriate ans\"atze for the metric and field
strength that enable us to study rather broad classes of cosmological
solutions, including in particular the ones utilising a 0-form field
strength, which we examine in section 3.

     In section 4, we examine the relationship between the
cosmological solutions obtained in the previous sections, and
non-extremal $p$-brane solitons.  In order to do this, it is necessary
to distinguish between two different kinds of generalisation of the
standard BPS-saturated extremal $p$-branes.  In the first of these,
which we shall refer to as type 1 non-extremal $p$-branes, the form of
the $D$-dimensional metric remains the same as in the extremal case,
namely $ds^2= e^{2A}\, dx^\mu\, dx_\mu + e^{2B}\, (dr^2 + r^2\,
d\Omega^2)$, where $A$ and $B$ are functions of $r$.  However, whereas
in the extremal case the function $X\equiv d A + \td d B$ is equal to
zero, here in the non-extremal generalisation it becomes non-vanishing
\cite{lpx1}.  (We have defined $d=p+1$ and $\td d=D-d-2$.)  The other
kind of non-extremal generalisation, which we shall refer to as type
2, begins from a modified form for the metric, namely $ds^2=e^{2A}\,
(-e^{2f}\, dt^2 + dx^i\, dx^i) + e^{2B}\, (e^{-2f}\, dr^2 + r^2\,
d\Omega^2)$ \cite{hs,dlp1}.  In this case, the relation $d A + \td d
B=0$ is still maintained.  Although both the type 1 and type 2
non-extremal generalisations introduce an additional function, namely
$X$ or $f$, the way in which they enter the metric ansatz is quite
different, although the two become equivalent when $p=0$.  As we shall
see, both types of non-extremal $p$-brane generalisation can be mapped
over into cosmological solutions.  The more standard cosmological
solutions actually correspond to the type 1 black $p$-branes; for this
reason, we shall refer to such cosmological solutions as type 1.
These are the kind that are constructed in sections 2 and 3.  We also
discuss some of the cosmological solutions associated with the type 2
black $p$-branes; some examples of this type have been considered in
\cite{lw}.  Finally in section 4, we also show how certain special
cases of the type 1 solutions can in fact be obtained by a dimensional
reduction of type 2 solutions, where the original time coordinate is
used for the compactification process.

\section{General cosmological solutions in string or M theories}

       The study of the cosmological consequences of string theory has been an
area of much active research in the past [7-11]. Recently, 
considerable attention has been directed to string-inspired cosmology
in various dimensions [12-22].  Cosmological models are described by
solutions of the low-energy effective theory in which the metric
tensor, and the other fields, are time dependent.  We shall consider
cosmological metrics of the form
\be
ds^2 = -e^{2U}\, dt^2 + e^{2A}\, d\bar s_q^2 + e^{2B}\, d\bar s_\q^2\ ,
\label{metric1}
\ee
where $U$, $A$ and $B$ are functions only of $t$, and $d\bar s_q^2$ and 
$d\bar s_\q^2$ denote metrics on maximally-symmetric spaces of positive,
negative or zero curvature, with dimensions $q$ and $\q$ respectively,
{\it i.e.}\
\be
d\bar s_q^2 = \fft{dr^2}{1-k r^2} + r^2 d\Omega^2\ ,\qquad
d\bar s_\q^2 = \fft{d\td r^2}{1-\td k \td r^2} + \td r^2 d\wtd \Omega^2\ ,
\ee
where $d\Omega^2$ and $d\wtd \Omega^2$ are the metrics on unit
$(q-1)$- and $(\q-1)$-spheres respectively.  For later convenience, we
define $d=q-1$ and $\td d = \q$ and hence $d+ \td d = D-2$.  The
constant $k$ and $\td k$ can be taken to be equal to 0, 1, or $-1$
independently, corresponding to flat, spherical or hyperboloidal
spatial sections respectively.  A large class of cosmological
solutions with $\td k=0$ were obtained in \cite{lmpx}.  The solutions
include some realistic models, where the $(q+1)$-dimensional spacetime
expands at large time, while the $\q$-dimensional space contracts and
becomes unobservable at large time. Note that when $\td k=0$, the
solution can be compactified on any $\q$-dimensional Ricci-flat space
with metric $ds_\q^2$.  In this section, we generalise the
previously-known solutions to include the cases where $\wtd k\ne 0$,
and briefly discuss their cosmological characteristics.

     In the vielbein basis $e^0=e^U\, dt$, $e^a=e^A\, \bar e^a$, $e^\a=e^B\,
\bar e^\a$, we find that the curvature 2-forms are given by
\bea
\Theta^0{}_a &=& e^{-2U}\, (\ddot A -\dot U \dot A + \dot A^2) e^0\wedge e^a
\ ,\nn\\
\Theta^0{}_\a &=& e^{-2U} \, (\ddot B -\dot U \dot B +\dot B^2) 
e^0\wedge e^\a
\ ,\nn\\
\Theta^a{}_b &=& \overline\Theta^a{}_b + e^{-2U} \dot A^2 e^a\wedge e^b\ ,
\label{2form}\\
\Theta^\a{}_\b &=&\overline\Theta^\a{}_\b + e^{-2U}\, \dot B^2 \, e^\a\wedge
e^\b\ ,\nn\\
\Theta^a{}_\b &=& e^{-2U}\, \dot A \dot B \, e^a\wedge e^\b\ ,\nn
\eea
where a dot denotes a derivative with respect to the time coordinate $t$, 
$\overline\Theta^a{}_b$ is the curvature 2-form for the metric $d\bar 
s_q^2=\bar e^a \bar e^a$ in the vielbein basis
$\bar e^a$, and  $\overline\Theta^\a{}_\b$ is the curvature 2-form
for the metric $d\bar  s_\q^2=\bar e^\a \bar e^\a$ in the vielbein basis
$\bar e^\a$.   It follows that the tangent-space components of the Ricci
tensor for the metric (\ref{metric1}) are given by 
\bea 
R_{00} &=& -e^{-2U}\Big(q ( \ddot A + {\dot A}^2 -\dot U \dot A ) 
+\q ( \ddot B + {\dot B}^2 - \dot U \dot B ) \Big) \ ,\nonumber\\
R_{ab} &=& e^{-2U} ( \ddot A + q {\dot A}^2 - \dot U \dot A 
+ \q \dot A \dot B ) {\delta}_{ab} + e^{-2A} \bar R_{ab} \ ,\label{ricci}\\
R_{\a\b} &=& e^{-2U} ( \ddot B + \q {\dot B}^2 - \dot U \dot B
+ q \dot A \dot B ) {\delta}_{\a\b}+ e^{-2B} \bar R_{\a\b} \ .\nonumber
\eea
where $\bar R_{ab}$ and $\bar R_{\a\b}$ denote the tangent-space components
of the Ricci tensor for the $q$-dimensional and $\q$-dimensional spatial
metrics. In all the cases we shall consider, these metrics will be
Einstein, and we may write $\bar R_{ab} = k(q-1)\delta_{ab}$ and
$\bar R_{\a\b} = \td k (\q-1)\delta_{\a\b}$. 

      We begin by solving for the simplest cosmological solutions in $D$
dimensions, which involve only the metric, a dilaton and an $n$-rank
antisymmetric field strength $F_n$.  The Lagrangian is given by
\be
e^{-1} {\cal L} = R - \ft12 (\del \phi)^2 - \fft1{2n!} e^{a\phi}
F_n^2\ ,\label{slag1}
\ee
where the constant $a$ can be conveniently parameterised as
\be
a^2 = \Delta -\fft{2(n-1)(D-n-1)}{D-2}\ ,\label{avalue}
\ee
and the constant $\Delta$ is preserved under the Kaluza-Klein
dimensional reduction \cite{lpss1}.   In supergravities, the full bosonic
Lagrangian can be consistently truncated to the single-scalar
Lagrangian (\ref{slag1}) for $\Delta=4/N$, where $N$ is a set of
integers $1, 2, \ldots, N_{\rm max}$, and $N_{\rm max}$  depends on
$D$ and $n$ \cite{lpsol}.
For 1-form field strengths, additional values of $\Delta$ can arise,
namely $\Delta = 24/(N(N+1)(N+2))$ \cite{lpsln}.  The equations of motion
from the Lagrangian (\ref{slag1}) are
\bea
\Box \phi &=& \fft{a}{2n!} e^{a\phi} F^2\ , \qquad
\del_{\sst M_1}( e e^{a\phi} F^{M_1\cdots M_n} ) = 0\ ,\nonumber\\
R_{\sst{MN}} &=& \ft12 \del_{\sst M} \phi \del_{\sst N} \phi +
S_{\sst{MN}}\ ,\label{seom}
\eea
where $S_{MN}$ is a symmetric tensor given by
\be
S_{\sst{MN}} = \fft{1}{2(n-1)!}\, e^{a\phi} \Big(F_{\sst{MN}}^2 -
\fft{n-1}{n(D-1)} F^2 g_{\sst{MN}}\Big)\ ,
\ee

    There are two types of ans\"atze for the field strength $F_n$ that are 
compatible with the symmetries of the metric (\ref{metric1}), giving 
rise to ``electric'' and ``magnetic'' cosmological solutions.  In the 
electric 
solutions, the ansatz for the antisymmetric tensor is given in terms of its 
potential, and in a coordinate frame takes the form
\be
A_{\a_1\a_2\cdots \a_\q} = f \epsilon_{\a_1\a_2\cdots \a_\q}\
,\label{eleans}
\ee
and hence
\be
F_{0\a_1\a_2\cdots \a_\q} = \dot f \epsilon_{\a_1\a_2 \cdots \a_\q}
\ ,\label{eleans1}
\ee
where $f$ is a function of $t$ only.  Here and throughout this paper 
$\epsilon_{\sst M\cdots \sst N}$ and $\epsilon^{\sst M \cdots \sst N}$
are taken to be tensor densities of weights $-1$ and $1$ 
respectively, with purely numerical components $\pm 1$ or $0$.  Note in 
particular that they are not related just by raising and lowering indices 
using the metric tensor.   For electric solutions, we have $q=D-n$ and 
$\q=n-1$.

      For the magnetic cosmological solutions, the ansatz for the 
tangent-space components for the antisymmetric tensor is
\be
F_{a_1 a_2\cdots a_q} = \lambda e^{-qA}\,\epsilon_{a_1a_2 \cdots a_q} \ ,
\label{solans}
\ee
where $\lambda$ is a constant.  Thus we have $q=n$ and $\q=D-n-1$.  The form
of the exponential prefactor is determined by the requirement that $F_n$
satisfy the Bianchi identity $d F_n=0$. 

     Substituting the ans\"atze for the metric and the field strength
into the equations of motion (\ref{seom}), we find
\bea
\ddot \phi + (q\dot A + \q\dot B -\dot U) \dot \phi &=& \ft12 \epsilon
a \lambda^2 \, e^{-\epsilon a\phi -2qA+2U}\ ,\nonumber\\
\ddot A + (q\dot A + \q\dot B - \dot U) \dot A + k(q-1) e^{2U-2A}&=&
\fft{\q}{2(D-2)} \lambda^2 \, e^{-\epsilon a\phi -2qA+2U}\ ,\nonumber\\
\ddot B + (q\dot A + \q\dot B - \dot U) \dot B + 
\tilde k (\q-1) e^{2U-2B}&=&
-\fft{q-1}{2(D-2)} \lambda^2 \, e^{-\epsilon a\phi -2qA+2U}\ ,\label{4eoms}\\
q(\ddot A + \dot A^2 - \dot U \dot A) +
\q(\ddot B + \dot B^2 - \dot U \dot B) + \ft12 \dot \phi^2 &=& 
-\fft{q-1}{2(D-2)} \lambda^2 \, e^{-\epsilon a\phi -2qA+2U}\ ,\nonumber
\eea
where $\epsilon =1$ for the electric case and $\epsilon=-1$ for the
magnetic case.  In the electric case, the constant $\lambda$ arises as 
the integration constant for the function $f$ in (\ref{eleans}).

      Let us first consider the equations of motion (\ref{4eoms})
for single-scalar cosmological solutions in $D$ dimensions.  It is
convenient to make the gauge choice $U=qA + \q B$, and to define
\be
Z_1 = A - \fft{\q}{\epsilon a (D-2)} \phi\ ,\qquad
Z_2 = B + \fft{q-1}{\epsilon a (D-2)} \phi\ .
\ee
The equations of motion for $Z_1$, $Z_2$ and $\phi$ become
\bea
\ddot Z_1 &=& -k(q-1) e^{2(q-1) Z_1 + 2 \q Z_2}\ ,\nonumber\\
\ddot Z_2 &=& -\td k (\q-1) e^{2q Z_1 + 2 (\q-1) Z_2 +
2\epsilon\phi/a} 
\ ,\label{todalike1}\\
\ddot \phi &=& \ft12\epsilon a \lambda^2 e^{2\q Z_2 -
\epsilon\Delta\phi/a}\ ,\nonumber
\eea
together with the first integral:
\bea
(2\q \dot Z_2 \!\!\!&-&\!\!\! \fft{\Delta}{\epsilon a} \dot \phi)^2 +
\Delta\lambda^2 e^{2\q Z_2 - \epsilon\Delta \phi/a} +
\fft{2\q(D-2)a^2}{q-1} \dot Z_2^2\nonumber\\
&=& \fft{2q\Delta}{q-1}((q-1) \dot Z_1 + \q \dot Z_2)^2 + 2\Delta
kq(q-1) e^{2(q-1) Z_1 + 2\q Z_2}\label{cons0}\\
&& + 2\Delta \td k \q(\q-1) e^{2q Z_1 +2(\q-1) Z_2 + 2\phi/(\epsilon a)}
\ .\nonumber
\eea
For generic values of $k$, $\td k$ and the charge parameter $\lambda$, the
equations of motion (\ref{todalike1}) cannot be solved in a closed form.
The following are special cases where solutions can be obtained:

\noindent
\underline{{\it Case 1}:}

       A special class of solutions were obtained in \cite{lmpx}, namely
those with $\td k=0$, but with $k$ arbitrary.  In this case, a further 
redefinition of fields was performed, namely
\be
X=\q Z_2 + (q-1) Z_1\ ,\qquad Y = Z_2\ ,\qquad \Phi =
-\fft{\Delta}{\epsilon a} \phi - 2\q Z_2\ .
\ee
These fields satisfy the following equations:
\be
\ddot X + k (q-1)^2 e^{2X} =0\ ,\qquad
\ddot \Phi + \ft12 \Delta \lambda^2 e^\Phi =0 \ ,\qquad
\ddot Y=0\ .
\ee
It follows from (\ref{cons0}) that the first integral corresponds to a
conserHamiltonian for the above equations, 
\be
\dot \Phi^2 + \Delta \lambda^2 e^\Phi +
\fft{2\q(D-2)a^2}{q-1} \dot Y^2 = \fft{2q\Delta}{q-1} (\dot X^2 +
k(q-1)^2 e^{2X})\ .\label{cons1}
\ee
Thus $X$ and $\Phi$ both satisfy Liouville equations.  The manifest
positivity of the left-hand side of (\ref{cons1}) (assuming $\Delta
>0$) shows that the Hamiltonian
$\dot X^2 + k(q-1)^2e^{2X}$ for $X$ must be positive, and hence the
appropriate form of the solution is
\bea
e^{-X} &=&\cases{\fft{q-1}{c}\, \cosh(ct +\delta)\ , & if $k=1$;\cr
\fft{q-1}{c}\, \sinh(ct +\delta)\ , & if $k=-1$;\cr}\nn\\
X&=& -ct-\delta\ ,\quad \hbox{if $k=0$}, \label{cases}
\eea
where $c$ and $\delta$ are constants. Note that in taking the square root of
$e^{2X}$, the positive root should be chosen in the expression for $e^{-X}$.
The Hamiltonian $\dot\Phi^2 + \Delta\lambda^2\, e^{\Phi}$ for $\Phi$ is also
manifestly positive, and so the solution can be written as
\be
e^{-\ft12\Phi}= \fft{\lambda \sqrt\Delta}{2\beta}\, \cosh(\beta t + \gamma)
\ ,\label{psol}
\ee
where $\beta$ and $\gamma$ are constants.  The solution for $Y$ may be taken
to be simply
\be
Y= -\mu t\ .\label{ysol}
\ee
The constraint (\ref{cons1}) therefore implies that
\be
\beta^2= \fft{q\Delta c^2-\q(D-2) a^2 \mu^2}{2(q-1)}\ .\label{beta}
\ee

     In terms of the original functions $A$, $B$ and $U$ appearing in the
metric (\ref{metric1}), and the dilaton $\phi$, the solution takes the form
\bea
e^{\ft{\Delta(D-2)}{2\q} A} &=& \fft{\lambda \sqrt\Delta}{2\beta} \,
\cosh(\beta t +\gamma)\,
e^{\ft{a^2(D-2)\mu t}{2(q-1)}}e^{\ft{\Delta(D-2)}{2\q(q-1)} X} \ ,\nn\\
e^{-\ft{\Delta (D-2)}{2(q-1)} B} &=& \fft{\lambda \sqrt\Delta}{2\beta} \,
\cosh(\beta t +\gamma)\, e^{\ft{a^2(D-2)\mu t}{2(q-1)}}
\ ,\label{sssol}\\
e^{\ft{\Delta}{2\epsilon a}\phi} &=& \fft{\lambda \sqrt\Delta}{2\beta} \,
\cosh(\beta t +\gamma)\, e^{-\mu q t}\ ,\nn
\eea
together with $U= q A + \q B$.  In a case where there is no dilaton, the
solutions for $A$ and $B$ are again given by (\ref{sssol}), with $\mu=0$.
If instead $\q=0$, we have from (\ref{avalue}) that $a^2=\Delta$; the
solution for $\phi$ follows from (\ref{sssol}) by setting $\q=0$, and $A$ is
given by $A=X/(q-1)$, with $X$ given by (\ref{cases}).   

     The solutions presented above are single-scalar solutions, involving a
single field strength carrying an electric or a magnetic charge.  In
fact, when $\td k=0$ one can also obtain general solutions for
multi-scalar, or dyonic cosmological models \cite{lmpx}.  We shall see
in section 4 that these solutions are closely related to static
$p$-brane solutions.

\noindent
\underline{{\it Case 2}:}

      In this case we have $\td k \ne 0$, but with $k=0$.  The
solution for the function $Z_1$ is easily found to be $Z_1=\mu t$.  
To simplify the equations of motion for $Z_2$ and $\phi$, we define
\be
u_1 = Z_2 + \fft{q\Delta}{b} Z_1\ ,\qquad
u_2 = -\fft{\phi}{\epsilon a} - \fft{2q\q}{b} Z_1
\ ,
\ee
which satisfy
\be
\ddot u_1 = -\td k (\q-1) e^{2(\q-1) u_1 - 2 u_2}\ ,\qquad
\ddot u_2 = - \ft12 \lambda^2 e^{2\q u_1 + \Delta u_2}\ ,\label{u12}
\ee
where $b=(\Delta +2) \q - \Delta$. If instead $\td k=0$, the equations
can be solved in general, reducing to the situation discussed in case
1.  If both $\td k$ and the charge parameter $\lambda$ are
non-vanishing, we cannot obtain the general solution to these
equations.  However, there exists a special solution where $\dot u_1$
and $\dot u_2$ are proportional to one another.  Thus we may make the
ansatz $u_2 = c_1 u_1 + c$, where $c$ and $c_1$ are constants.  The
two equations (\ref{u12}) reduce to a single Liouville equation, if
$c_1=-2/(\Delta +2)$ and $e^{(\Delta +2)c} = - 4\td k(\q-1)/(\lambda^2
(\Delta +2))$. Defining $u=b/(\Delta+2) u_1$, this Liouville equation
is
\be
\ddot u = \ft14 b\lambda^2 \, e^{\Delta c} e^{2 u}
\ .\label{liouvilleu}
\ee 
The first order equation (\ref{cons0}) becomes
\be
\dot u^2 -\ft14 b\lambda^2 \,
e^{2u+\Delta c} = \fft{q(D-2) a^2\,\mu^2 }{2+ (\Delta +2) \q}\ .\label{cons2}
\ee
The Liouville equation (\ref{liouvilleu}) with the first-order
constraint (\ref{cons2}) can be easily solved, giving finally
\bea
e^{-u} &=& \fft{\lambda\sqrt{b} }{2\beta}\,
e^{\ft{\Delta}{2} c} \sinh(\beta t + \a)\ ,\label{case2sol}\\
Z_1&=&\mu t\ ,\quad u_1 = \fft{\Delta +2}{b} 
u\ ,\quad u_2 = -\fft{2}{b} u + c\ ,\nonumber
\eea
where $c$ is given above equation (\ref{liouvilleu}).  It follows from
(\ref{cons2}) that the constant $\beta$ is given by
\be
\beta^2 = \fft{q(D-2)a^2\, \mu^2}{(\Delta +2) \q + 2} \ .
\ee
In terms of the original functions $A$, $B$ and $U$ appearing in
the metric (\ref{metric1}), the solution is
given by
\bea
e^{-\ft{\Delta' (D-2)}{2\q} A} &=& \fft{\lambda\sqrt{b}}{2\beta} 
e^{\ft12\q(\Delta +2) c} \sinh(\beta t+\a) e^{(q\q -
\ft{b(D-2)}{2\q}) \mu t}\ ,\nonumber\\
e^{-\ft{b}{\Delta'+2} B} &=& \fft{\lambda \sqrt{b}}{2\beta}
\sinh(\beta t +\a) e^{(\ft12\Delta -\ft{(q-1)b}{(D-2)(
\Delta' +2)}) c} e^{\ft{q a^2}{\Delta' +2} \mu t}\ ,
\label{case2sol1}\\
e^{-\ft{b}{2\epsilon a} \phi} &=& \fft{\lambda\sqrt{b}}{2\beta}
e^{\ft12 \q(\Delta +2) c} \sinh(\beta t +\a) e^{q\q\mu t}\ ,\nonumber
\eea
where $\Delta'=\Delta -2(q-1)/(D-2)$.  

     It is of interest to examine the cosmological features of these
solutions. Note that unlike the $\td k=0$ case, there exists a limit
where $\beta =0=\mu$.  In this limit, all the functions $A$, $B$ and
$\phi$ depend linearly on $\log t$.  When $\beta$ and $\mu$ are
non-vanishing, the time dependence of these functions is more
complicated.  Let us consider $\beta$ and $\mu$ to be both positive.  At
the beginning, when $\sinh(\beta t + \a)$ vanishes, both the scale
factors $e^A$ and $e^B$ diverge.  As the coordinate $t$ approaches
infinity, $e^A$ and $e^B$ behave quite differently.  For $\Delta >0$,
the scale factor $e^{B}$ shrinks to zero, whilst the behaviour of the
scale factor $e^A$ depends on the sign of $\beta + \mu (q\q -
b(D-2)/2\q)$.  This can be shown to be negative if $a^2\q(\q-1) >
2$. The scale factor $e^A$ thus starts from infinity at the beginning and
shrinks to a minimum and then grows indefinitely at late times.  The
behaviour of the dilaton depends on the sign of $\epsilon$.

\noindent
\underline{{\it Case 3}:}

     In this case, we take both $k$ and $\td k$ to be non-vanishing.
As we mentioned earlier, the equations (\ref{todalike1}) cannot be
solved in general. However there exists a special solution, where the
first derivatives of the functions $Z_1$, $Z_2$ and $\phi$ are
proportional to one another.  The equations (\ref{todalike1}) can be
reduced to a single Liouville equation if we have
\bea
Z_1 &=& \fft{\Delta}{(D-2) a^2} u + \fft{(\Delta - \Delta \q -\q) x_1 +
\Delta \q x_2 + 2 \q x_3}{2(D-2) a^2}\ ,\nonumber\\
Z_2 &=& \fft{\Delta +2 - 2q}{(D-2) a^2} u+ \fft{\Delta q x_1 - \Delta
(q-1) x_2 - 2(q-1) x_3}{2(D-2) a^2}\ ,\\
-\fft{\phi}{\epsilon a} &=& \fft{2(q-1)}{(D-2) a^2} u + 
\fft{q\q x_1 - (q-1) \q x_2 - (D-2)x_3}{(D-2) a^2}\ ,\nonumber
\eea
where the constants $x_1$, $x_2$ and $x_3$ are given by
\be
e^{x_1} = \fft{\Delta}{(-k)(D-2) (q-1) a^2} \ ,\quad
e^{x_2} = \fft{\Delta +2 -2q}{(-\td k)(D-2) (\q-1) a^2}\ ,\quad
e^{x_3} = \fft{4(q-1)}{\lambda^2 (D-2) a^2}\ .\label{relat}
\ee 
The Liouville equation is given by $\ddot u - e^{2u}=0$, and the
first-order equation becomes $\dot u^2 - e^{2u} =0$, with the solution
$u = -{\rm log}\,t + c$. Here $c$ is a constant of integration. 
Thus using the freedom to choose $c$, we can relax one of the
relations in (\ref{relat}). In this case, the behaviour of the metric
components is
\be
e^A \, \sim \, t^{-{1\over D-2}},\quad e^B \,\sim \, 
t^{-{1\over D-2}},\quad e^{{\phi\over {\epsilon a}}}\,
\sim \,t^{{2(q-1)\over {(D-2)a^2}}},
\ee
and so $e^A$ and $e^B$ decrease as $t$ increases.  For electric
solutions $(\epsilon = 1)$ the dilaton $\phi$
tends to infinity for large $t$, while for magnetic solutions $(\epsilon
= -1)$ the dilaton tends $-\infty$ as $t$ increases.
It is also not hard to express the scale factors in terms
of the comoving time $\tau$:
\be
e^A \, \sim \, \tau, \quad e^B \, \sim \, \tau \quad
e^{\phi\over{\epsilon a}} \, \sim \, \tau^{2(q-1)\over a^2}.
\ee

\section{Cosmological solutions with cosmological terms}

      In the previous section, we obtained cosmological solutions
using $n$-form field strengths.  Such solutions arise in massless
supergravities in $D$-dimensions.  Supergravities can also admit one
or more cosmological terms, which can be viewed as the special case
where $n=0$.  Such theories are in general massive, but they can
nevertheless be obtained by the Kaluza-Klein reduction of massless
supergravities in higher dimensions.  For example, gauged massive
supergravity in $D=4$ \cite{dn} or $D=7$ \cite{tv} can be obtained by
the dimensional reduction of 11-dimensional supergravity \cite{cjs} on
a 7-sphere \cite{dp} or a 4-sphere \cite{pnt}.  The highest dimensional
massive supergravity is the massive type IIA theory in $D=10$
\cite{r}, which seems to indicate the possible existence of a
13-dimensional theory \cite{clpst}.  In fact, the number of massive
supergravities in lower dimensions is far greater than that of
massless supergravities, at least in the maximally supersymmetric
case.  It was shown recently that certain generalised Kaluza-Klein
compactifications of the low-energy limits of string theory or
M-theory can give rise to massive supergravities with cosmological
terms in $D$-dimensions \cite{bdgpt,clpst,lpdomain,llp}.  The
generalised reduction involves making an ansatz for a rank $(n-1)$
potential such that its field strength contains a term that is a
constant multiple of an harmonic $n$-form on the compactifying
manifold.  Such a phenomenon was also discussed earlier from a group
theoretic point of view in \cite{ss}.  In general, such massive
supergravities contain cosmological terms.  We are interested in cases
where the bosonic Lagrangian can be consistently truncated to one
containing just the metric, dilatonic scalar fields and cosmological
terms, of the form:
\be
e^{-1} {\cal L} = R - \ft12 (\del \vec \phi)^2 -
                \ft12 \sum_\a^{N} m_\a^2\,e^{\vec c_\a \cdot \vec \phi}
\ ,\label{multilag}
\ee
where $\vec \phi=(\phi_1, \phi_2, \ldots )$ denotes a set of dilatonic
scalar fields in the theory and $\vec c_\a$ are constant vectors.
In the toroidal compactification, consistency of the truncation
requires that the dilatonic vectors $\vec c_\a$ satisfy the
following dot product relations  \cite{lpsol}
\be
M_{\a\beta} = 4 \delta_{\a\beta} + \fft{2(D-1)}{D-2}\ .\label{mmatrix}
\ee
In this case, the Lagrangian (\ref{multilag}) can be further truncated
to a single-scalar Lagrangian, of the form \cite{lpsol}
\be
e^{-1}{\cal L} = R - \ft12 (\del \phi)^2 -\ft12 m^2 e^{a\phi}
\ ,\label{slag2}
\ee
where 
\bea
a^2 &=& \Big(\sum_{\a,\beta}(M^{-1})_{\a\beta} \Big)^{-1}\ ,\qquad
\phi = a\sum_{\a,\beta} (M^{-1})_{\a\beta}\, \vec c_\a\cdot \vec \phi\ ,
\nonumber\\
m_\a^2 &=& a^2\, m^2\, \sum_{\beta} (M^{-1})_{\a\beta} = \fft{m^2}{N}\ .
\eea

     The constant $a$ can be parameterised by writing $a^2=\Delta +
2(D-1)/(D-2)$, with $\Delta = 4/N$. In some cases, the Lagrangian
(\ref{slag2}) can be embedded in a massive supergravity arising from
the generalised reduction of the low-energy effective action of a
string theory or M-theory on some other Ricci-flat manifold, such as
K3, Calabi-Yau or a 7-dimensional Joyce manifold \cite{lpdomain,llp}.
The Lagrangian (\ref{multilag}) gives rise to supersymmetric
domain-wall solutions, which can be oxidised back to $D=10$ or $D=11$
where they become $p$-branes or intersecting $p$-branes.  The
situation for compactifications of string theory or M-theory
where the internal manifold is not Ricci-flat is different.
In particular, values of $\Delta$ other than $4/N$ can now arise in
the resulting massive supergravity theories in the lower dimensions.
For example, in gauged massive supergravity in $D=7$, which can be
obtained by compactifying 11-dimensional supergravity on a 4-sphere,
the value of $\Delta$ for a certain cosmological term is $-2$.  The
corresponding domain-wall solutions, although they can still be
oxidised to $D=11$ owing to the consistency of the reduction, will no
longer become $p$-branes or intersecting $p$-branes.  For example a
special domain-wall solution, namely the AdS$_7$ spacetime solution of
7-dimensional gauged massive supergravity, becomes AdS$_7\times S^4$
upon oxidation, and thus cannot be regarded as a $p$-brane in $D=11$.
In other words, in the generalised dimensional reduction discussed
above, the harmonic function associated with the domain-wall solution
in the lower dimension is precisely the same as the harmonic function
associated with the higher-dimensional $p$-brane solution, implying
that the lower-dimensional solution can be obtained from
higher-dimensional one {\it via} the vertical dimensional reduction
procedure \cite{kh,ghl,ht,lps}.  By contrast, in the spherical
compactification the harmonic function in the lower-dimensional
domain-wall solution is not the same as the one in the
higher-dimensional solution.

     In this section, we are focussing on the construction of
cosmological solutions that involve cosmological terms.  Such
solutions are magnetic, in the sense that they are the extrapolation
to $n=0$ of the magnetic ansatz discussed in section 2.  Of course
the cosmological terms can be dualised to give rise to $D$-form field
strengths in $D$ dimensions.  In terms of the $D$-form field
strengths, the corresponding solutions are instead electric.  The
cosmological solutions that we are going to discuss include both the
magnetic solutions for cosmological terms and the electric solutions
for $D$-form field strengths.  It follows from (\ref{metric1}) and the
fact that $q=0$ and $\q=D-1$ for both cases that the metric ansatz is
given by
\be
ds^2 = -e^{2(D-1) B} dt^2 + e^{2B} ds_\q^2\ ,\label{metric2}
\ee
where we have chosen the gauge $U=(D-1) B$.  We shall first consider
solutions from the single-scalar Lagrangian (\ref{slag2}) (or its
$D$-form dual with $e^{-a\phi}\, F_{\sst D}^2$).  The equations of
motion are given by
\bea
&&\ddot \phi = -\ft12 a m^2\, e^{2(D-1) Y + \Delta \phi
/a}\ ,\qquad \ddot Y = - \td k (D-2) e^{2(D-2) Y - 
2\phi/a}\ ,\nonumber\\
&&\ft12 m^2\, e^{2(D-1) Y +\Delta \phi/a} -
(D-1)(D-2)(\dot Y + \fft{\dot\phi}{(D-2)\epsilon a})^2 +
\ft12 \dot \phi^2 \\
&&= \td k (D-1) (D-2) e^{2(D-1) Y - 2\phi/a}\ .\nonumber
\eea
These equations can be solved in general when $\td k=0$, giving a
special case of the $\td k=0$ solutions discussed in section 2.
The solution is
\bea
e^{\ft{\Delta(D-2)}{2} B} &=& \fft{m\sqrt{\Delta}}{2\beta}\,
\cosh(\beta t +\a) e^{\ft{(D-2) a^2}{2} \mu t}\ ,\nonumber\\
e^{-\ft{\Delta\phi}{2a}} &=& \fft{m\sqrt{\Delta}}{2\beta}\,
\cosh(\beta t + \a) e^{(D-1)\mu t}\ ,\label{cosmosol1}
\eea
where the constants $\beta$ and $\mu$ are related by 
\be
\beta^2 =\ft12 (D-1)(D-2) a^2\mu^2\ .\label{beta1}
\ee
    
          In this $\td k=0$ case, the generalisation
of the solution (\ref{cosmosol1}) to a multi-scalar solution
with the dilaton vectors $\vec c_\a$ satisfying the
dot products (\ref{mmatrix}) is straightforward, and the solution is
given by
\bea
e^{2(D-2) B} &=& e^{\ft{2(D-2)a^2}{\Delta} \mu t} \prod_{\a=1}^{N}
\Big(\fft{m_a}{\beta_\a} \cosh(\beta_\a t + \gamma_\a)\Big)
\ ,\nonumber\\
e^{-\ft12 \vec c_\a \cdot \phi - (D-1) B} &=&
\fft{m_\a}{\beta_\a} \cosh(\beta_\a + \gamma_\a)\ ,
\eea
where $\Delta = 4/N$ and the constants $\beta_a$ and $\mu$ are related
by $\sum_\a \beta_\a^2 = 2(D-1)(D-2) a^2\mu^2/\Delta$.

     Having obtained cosmological solutions involving cosmological
terms, it is of interest to study their physical characteristics.  For
now, we shall consider them from the point of view of the
$D$-dimensional theory itself.  The Lagrangian (\ref{slag2}) has a
coupling constant given by $g_{\rm m} = e^{-\ft12 a\phi}$.  (The
coupling constant for the dual theory with a $D$-form field strength
is given by $g_{\rm e} = 1/g_{\rm m}$, since the dilaton reverses its
sign under the dualisation.)  In order to study the large $|t|$
behaviour of the coupling constant, we see from (\ref{cosmosol1}) that
we need to compare the values of $\beta$ and $(D-1)\mu$:
\be
\beta^2 - (D-1)^2 \mu^2 = \ft12 (D-1)(D-2)\Delta \mu^2\ .
\ee
Thus the behaviour of the coupling constant depends on the sign of the
constant $\Delta$.  For positive $\Delta$, the coupling constant
$g_{\rm e}$ tends to zero at large $|t|$, whilst the coupling constant
$g_{\rm m}$ tends to infinity.  For cosmological terms in supergravity
theories, the $\Delta$ values can sometimes be negative instead.  For
example, a pure cosmological term without any dilaton coupling has
$\Delta =-2(D-1)/(D-2)$. (In this case, however, the discussion for
the dilaton behaviour is irrelevant.)  Another example is provided by
gauged massive supergravities in $D=4$ or $D=7$, which arise from the
spherical compactification of 11-dimensional supergravity.  The value
of $\Delta$ for the cosmological term associated with the gauge
parameter is $\Delta =-2$.  In this case, the coupling constant
$g_{\rm e}$ runs from infinity to zero as the coordinate $t$ goes from
$-\infty$ to $\infty$, and the coupling constant $g_{\rm m}$ behaves
in exactly the opposite way (here without loss of generality we assume
that $\mu$ is positive).  Similarly, we can discuss the behaviour at
large $|t|$ for the scale factor $R=e^B$.  In this case, we need to
compare the value of $\beta$ with $a^2(D-2)/2\mu$:
\be
\beta^2 - (\ft12(D-2) a^2\mu)^2 = -\ft14 (D-2)^2 a^2\Delta\ \mu^2.
\ee
Again the behaviour of the scale factor $R$ depends only on the
sign of $\Delta$.  If $\Delta$ is negative, the scale factor
diverges at large $|t|$.  On the other hand, if $\Delta$ is positive,
the scale factor runs from zero to infinity as the $t$ coordinate runs
from $\infty$ to $-\infty$.

      In order to discuss the evolution of the solutions, it is useful
to introduce the comoving time coordinate $\tau$, defined by $ds^2 =
-d\tau^2 + e^{2B} ds_\q^2$, which implies that $\tau = \int^t
e^{(D-1)B} dt$.  Thus at large $t$, we have $\tau \sim
\exp((D-1)(\beta |t| - \ft12 (D-2) a^2 t) /(2(D-2))) $.  If $\Delta$
is positive, the universe starts at finite $\tau_0$ when the scale
factor is zero, and then expands to infinity as $\tau$ goes to
infinity. In this case, since we have $R=\tau^{1/(D-1)}$, the
expansion of the universe slows down at large $\tau$, but there is no
inflation in the early stages.  If $\Delta$ is negative, the situation
is different.  In this case, the comoving time $\tau$ runs from
infinity to a finite value and then to infinity again as $t$ goes from
$-\infty$ to $\infty$. Thus we can also consider that the universe
starts at a finite $\tau_0$ when its size is minimal, and then it
expands to infinity as $\tau$ goes to $\infty$.

       The physical properties of the cosmological solutions with
cosmological terms were discussed above in the context in the
$D$-dimensional theories themselves.  Since these theories are
dimensional reductions of more fundamental theories such as
string theory or M-theory, their cosmological features can be more
appropriately analysed in ten or eleven dimensions.  As we mentioned
earlier, there are numerous massive supergravity theories in lower
dimensions, and we shall give only a few examples to illustrate the
oxidation of the lower-dimensional solutions and to discuss their
higher-dimensional characteristics.

       One simple example is the electrically-charged cosmological
solution in $D=4$ using the 4-form coming from the
4-form field strength in M-theory in $D=11$.  The 4-dimensional
solution is given by
\bea
e^{4B} &=& \fft{m}{\beta} \cosh(\beta t +\a) e^{7\mu t}\ ,\nonumber\\
e^{-2\phi/\sqrt7} &=&\fft{m}{\beta} \cosh(\beta t + \a) e^{3\mu t}\ ,
\label{aahh}
\eea
where $\beta =\sqrt{21} \mu > 0$.   The solution can be oxidised to
$D=11$, giving
\bea
ds_{11}^2 &=& e^{\ft{\sqrt 7}{3}\phi} ds_4^2 +
e^{-\ft{2}{3\sqrt7}\phi} ds_7^2
\nonumber\\
&=& -e^{6B + \ft{\sqrt7}{3} \phi} dt^2 + e^{2B + \ft{\sqrt7}{3} \phi} 
ds_3^2 +  e^{-\ft{2}{3\sqrt 7}\phi} ds_7^2\ .\label{d11sol1}
\eea
In this 11-dimensional metric, we see that the scale factor for the
3-dimensional space $R=e^{B + \phi\sqrt7/6} \sim (\cosh(\beta t +
\a))^{-1/2}$ now shrinks to zero at large $|t|$.  On the other hand,
the scale factor $R'=e^{-\phi/(3\sqrt7)}$ for the 7-dimensional
internal space runs from infinity to a minium size and then to
infinity again as the coordinate $t$ runs from $-\infty$ to $\infty$.
Note that the comoving time coordinate $\tau$ runs from 0 to $\tau_0$,
where $R'$ reaches its minimum, and then goes to infinity.  Thus the
11-dimensional metric effectively describes an 8-dimensional expanding
universe which starts at time $\tau_0$, where it has a Plankian size.
Note that the time interval $(0, \tau_0)$ is of the Plankian scale
too.  Since the scale factor $R'$ has only one minimum, at
$\tau=\tau_0$, it implies that $\dot R'\equiv dR'/d\tau > 0$ for all
times $ \tau_0 < \tau < \infty$.  It also implies that close to the
beginning of the universe, $\tau\sim\tau_0$, we have $\ddot R'>0$ and
hence an inflationary expansion.  On the other hand as
$\tau\rightarrow \infty$ we have $R'\sim \tau^{(\beta + 3\mu)/(\beta +
21\mu)}$, and hence $\ddot R'<0$, implying a slowing down of the
expansion rate of the universe.
 
     Now let us consider a 4-dimensional magnetic cosmological
solution in $D=4$ using a cosmological term.  The 4-dimensional theory
in question is obtained \cite{llp} from a generalised compactification
of M-theory on a 7-torus or on a 7-dimensional Joyce manifold
\cite{j}.  The solution involves 7 cosmological terms, and hence has
$a^2 = 25/7$ and $\Delta =4/7$ (see \cite{llp} for the configuration
of the cosmological terms).  We consider only the case where all the
cosmological constants are equal, for which the solution is given by
\bea
e^{\ft47 B} &=& \fft{m}{\sqrt7\beta} \cosh(\beta t+\a)
e^{\ft{25}{7}\mu t}\ ,\nonumber\\
e^{-\ft2{5\sqrt7} \phi} &=&  \fft{m}{\sqrt7\beta} \cosh(\beta t+\a)
e^{3\mu t}\ ,\label{bphi2}
\eea
where $\beta = 5\sqrt{3/7} \mu$.  Again this solution can be oxidised
to $D=11$, in fact giving the same result (\ref{d11sol1}) as in the
electric solution described above, except that now $B$ and $\phi$ are
given by (\ref{bphi2}) rather than (\ref{aahh}).  In this case, one
can easily verify that $R$ shrinks and $R'$ expands, effectively
describing an 8-dimensional expanding universe.  As in the example
above, at the beginning of the universe $\tau=\tau_0$ there is an
inflationary period, whilst at large times the expansion rate slows
done.

        An interesting massive supergravity theory is the one in $D=7$
with a topological mass term, which can be obtained by a generalised
compactification of M-theory on a 4-torus or a K3 manifold.  Here, the
generalised Kaluza-Klein ansatz implies that the 4-form field strength
has an extra term that is proportional to the volume form of the
internal space.  The cosmological constant in $D=7$ is the square of
this constant of proportionality.  The 7-dimensional solution
takes the form
\bea
e^{10B} &=& \fft{m}{\beta} \cosh(\beta t + \a) e^{16\mu t}
\ ,\nonumber\\
e^{-\ft12\sqrt{\ft52}\phi} &=& \fft{m}{\beta} \cosh(\beta t + \a) 
e^{6\mu t}\ ,
\eea
where $\beta = 4\sqrt{6}\mu$.  It can be oxidised to the $D=11$ solution
\bea
ds_{11}^2 &=& e^{\ft23\sqrt{\ft25}\phi} ds_7^2 +
e^{-\ft13\sqrt{\ft52}\phi} ds_4^2\nonumber\\
&=&- e^{12B + \ft23\sqrt{\ft25}\phi} dt^2 + e^{2B
+\ft23\sqrt{\ft25}\phi} ds_6^2
             + e^{-\ft13\sqrt{\ft52}\phi} ds_4^2\ ,
\eea
It is easy to see that now the scale factor for the six-dimensional
space with metric $ds_6^2$ is $R=e^{B + \ft13\sqrt{\ft25} \phi} \sim
(\cosh(\beta t + \a))^{-1/6}$, implying that the space shrinks to zero
at large $|t|$, whilst the scale factor $R'=e^{-\ft16\sqrt{\ft52}\phi}$
diverges at large $|t|$.  Thus the solution effectively describes a
5-dimensional expanding universe.  The comoving time coordinate runs
from zero to a Plankian time $\tau_0$, where the scale factor $R'$ is 
a minimum, and
then runs to infinity as the 4-dimensional space expands while the
6-dimensional space contracts.  Since at the starting point of the
universe $\tau=\tau_0$ the scale factor $R'$ is a minimum, we have
that $\ddot R'>0$
in the early universe. At large $\tau$, we have $R'\sim \tau^{(\beta +
6 \mu)/(\beta + 24 \mu)}$, implying that $\dot R'>0$ but $\ddot R' <
0$.

      We can alternatively oxidise the solution to $D=10$ instead. In
this case, it will effectively describe a 4-dimensional expanding
universe.  To see this, we note that the metric in $D=10$ is given by
\bea
ds_{10}^2 &=& e^{\fft{9}{8\sqrt{10}} \phi} ds_7^2 + 
e^{-\ft38\sqrt{\ft52}\phi} ds_3^2\nonumber\\
&=& -e^{12B + \ft{9}{8\sqrt{10}}}\, dt^2 +
e^{2B + \ft{9}{8\sqrt{10}}\phi} ds_6^2 + 
e^{-\ft38\sqrt{\ft52}\phi} ds_3^2\ .
\eea
Again the scale factor for the six-dimensional space with metric
$ds_6^2$ shrinks whilst the 3-dimensional space with metric $ds_3^2$
expands, effectively describing a 4-dimensional expanding universe,
with an inflationary period at early times, and a slowing down of the
expansion rate these days.  In fact this is a solution in $D=10$ which
is supported by a 3-form field strength.  General cosmological
solutions involving 3-form field strengths in $D=10$ were also
discussed in \cite{low1,lmpx}.  It is interesting to study the
behaviour of the string coupling in $D=10$.  There are two maximally
supersymmetric string theories in $D=10$, namely the type IIA and type
IIB strings.  If we oxidise the $D=7$ solution to $D=10$, where the
$D=7$ theory is taken to be embedded in the type IIA theory, we find
that the corresponding string coupling constant is given by
$g=e^{-\sqrt{5/2}\phi/4}$, which diverges at large $|t|$.  If on the
other hand we view the $D=7$ theory as being embedded in the type IIB
string, and oxidise it to $D=10$, there are two possible outcomes.  If
the charge is carried by the NS-NS 3-form, then the coupling constant
is the same of the type IIA case, and diverges at large $|t|$; on the
other hand, if the charge is carried by the R-R 3-form, then the
coupling constant is the inverse of that in the type IIA case, and
hence goes to zero at large $|t|$.

\section{Relation between cosmological and $p$-brane solutions}

    It has been observed recently that there is a similarity in
certain cases between the cosmological solutions of the kind we are
discussing in this paper, and type 1 black $p$-brane solutions in
string theory \cite{low1}.  In this section, we shall explore this in
a more general setting, and show that there is an exact one-to-one map
between the $k=1$, $\tilde k=0$ type 1 cosmological solutions and the
type 1 black $p$-brane solutions that were discussed in \cite{lpx1}.
In the special case when $p=0$, these latter solutions are the same as
standard non-extremal black holes.  Thus a subset of the type 1
cosmological solutions can be mapped into standard black-hole
solutions.  However, when $p$ is greater than zero, the type 1 black
$p$-brane solutions obtained in \cite{lpx1} are not ordinary type 2
\cite{dlp1} black $p$-branes.  Thus in cases where the cosmological
solutions map into $p$-brane solutions with $p>0$, it is to these
non-standard type 1 black $p$-branes, rather than the standard type 2
ones, that the mapping occurs.

     To see this in detail, we begin by recalling the structure of the
type 1 black $p$-branes constructed in \cite{lpx1}.  They have metrics
which take the form
\be
ds^2 = e^{2\wtd A}\, (-d\tilde t^2 + dx^i\, dx^i) + e^{2\wtd B}\, 
(d\td r^2 + \td r^2 d\Omega^2) \ ,\label{bpbrane}
\ee
where $\wtd A$ and $\wtd B$ are functions only of $\td r$.  Here $x^i$
are the $p$ spatial coordinates on the world-volume of the $p$-brane,
and $d\Omega^2$ is the metric on the unit $(\td d +1)$-sphere, where
$\td d=D-d-2$ and $d=p+1$.  The metric (\ref{bpbrane}) is of the form
of the standard ansatz for extremal $p$-branes, except that there the
further restriction $d \wtd A + \td d \wtd B=0$ is
imposed.  In \cite{lpx1}, more general solutions, which are
non-extremal, were obtained by relaxing this assumption, so that
there is an additional variable $\wtd X \equiv d \wtd A + \td d \wtd
B$ appearing in the equations of motion.  This has the solution
$e^{\wtd X} = 1- \td\kappa^2\rho^2$, where $\td\kappa$ is a constant
and $\rho\equiv \td r^{-\td d}$.  It is natural to introduce a further
redefined radial coordinate $\xi$, such that $e^X \del/\del\rho
=\del/\del\xi$, and thus $\td\kappa\rho =\tanh \td\kappa\xi$.  The
general type 1 black $p$-brane solutions take the form \cite{lpx1}
\bea
e^{-\ft{(D-2)\Delta}{2\td d} \wtd A} &=& 
\fft{\td \lambda\sqrt{\Delta} }{2\td d \td \beta}\, \sinh(\td \beta \xi +\a) 
\, e^{a^2 (D-2) \td\mu \xi/(2\td d)}\ ,\nn\\
e^{\ft{(D-2)\Delta}{2 d} \wtd B} &=& 
\fft{\td \lambda\sqrt{\Delta} }{2\td d \td \beta}\, \sinh(\td \beta \xi +\a) 
\, e^{a^2 (D-2) \td\mu \xi/(2\td d)}\, 
(\cosh \td\kappa\xi)^{-\ft{(D-2)\Delta}{d \td d}}\ ,\label{bpsol}\\
e^{\ft{\Delta}{2a\epsilon}\phi} &=&
\fft{\td \lambda}{2\td d \td \beta}\, \sinh(\td \beta \xi +\a)
\, e^{-d\td\mu\xi}\ .\nn
\eea
The various constants of integration are subject to the constraint 
$\td d \td\beta^2 = 2(\td d+1) \Delta \td\kappa^2 -\ft12 
a^2 d(D-2)\td\mu^2$.

     The metrics described by these solutions have an outer event
horizon where the function $e^X=1-\td\kappa^2\rho^2$ vanishes.  In the
case when $p=0$, they describe precisely the standard non-extremal
black-hole solutions of string theory, with $\td\kappa$ being a
non-extremality parameter that vanishes in the extremal limit.
However, when $p$ is greater than zero, (\ref{bpbrane}) is quite
different from the metric of a standard type 2 black $p$-brane, for
which one would have \cite{dlp1}
\be
ds^2= e^{2\hat A} (-e^{2f} \, dt^2 + dx^i \, dx^i) + e^{2\hat B} (e^{-2f} \, 
dr^2 + r^2\, d\Omega^2) \ ,\label{stand}
\ee
with $e^{2f}= 1-\kappa r^{-\td d}$ and
\bea
e^{-\ft{(D-2)\Delta}{2\td d} \hat A} &=& 1+ \fft{\kappa}{r^{\td d}}\, 
\sinh^2\mu\ ,\nn\\
e^{\ft{(D-2)\Delta}{2 d} \hat B} &=& 1+ \fft{\kappa}{r^{\td d}}\, 
\sinh^2\mu\ ,\label{standsol}\\
e^{\ft{\Delta}{2a\epsilon}\phi} &=& 1+ \fft{\kappa}{r^{\td d}}\, 
\sinh^2\mu\ .\nonumber
\eea
These solutions also have an outer horizon where $e^{2f}=1-\kappa
r^{-\td d}$ vanishes.  However, it is clear that the structure of
these metrics is not the same as those of (\ref{bpbrane}) when $p>0$.
In particular, in (\ref{bpbrane}) the sign of the spatial $p$-brane
metric $dx^i \, dx^i$, as well as the $dt^2$ part, also reverses when
the horizon is crossed, whereas in the standard metrics (\ref{stand})
only the $dt^2$ part of the $p$-brane world-volume metric reverses
sign.  The two types of solution become equivalent if $p=0$.

     It is clear that the solutions (\ref{bpsol}) have a very similar
structure to the cosmological solutions given in section 2.  In fact,
as we shall show below, the type 1 $p$-brane solutions (\ref{bpbrane})
and (\ref{bpsol}) can be mapped into the type 1 cosmological solutions
(\ref{sssol}) with $k=1$ and $\td k=0$.  Under this mapping, the
$q$-dimensional space in the cosmological solution, which is flat
since $\td k=0$, becomes the world volume of the $p$-brane, with
$p=\q-1$.  Depending on whether the function $e^{2B}$ in
(\ref{metric1}) is positive or negative after the mapping, either one
or $(\q-1)$ of these Euclidean coordinates must be Wick rotated.  The
coordinate $t$, which becomes spacelike either through a Wick rotation
or because $e^{2U}$ becomes negative, together with the coordinates of
the $q$-dimensional space in (\ref{metric1}), become the coordinates
of the transverse space of the $p$-brane solution.

     To see this, it is instructive first to write the metric 
(\ref{metric1}) for $k =1$ and  $\tilde k =0$ in the following form:
\be
ds^2 = -e^{2U}\, dt^2 + e^{2A}\, d\bar s_q^2 + e^{2B}\, (dy^2
+ dy^j \, dy^j)\ ,\label{abcmetric}
\ee
where $j$ runs over $\q-1$ values.
It is clear that the $p$-brane solutions can be 
mapped to the cosmological ones if the following relations hold:
\be 
e^{2{\wtd B}} dr^2 = -e^{2U} dt^2, \quad e^{2\wtd B}r^2 = e^{2A},
\quad e^{2\wtd A} = (-1)^\delta e^{2B}\ ,
\ee
where $\delta$ takes the value $0$ or $1$. Furthermore, we need to 
identify $\tilde t = i y,  \, x^i = y^j$ if $\delta =0$ or 
$\tilde t = y, \, 
x^j = \im y^j$ if $\delta =1$, implying $d = \q$ and 
$\tilde  d = q-1$. From the above relations, we immediately see 
that the $p$-brane coordinate $\xi$ and the constant $\kappa$ are 
related to the time coordinate $t$ and the constant $c$ of the 
cosmological solutions by 
\be
\xi = i^{-\q\delta + 1}(q-1)t + \nu, \qquad \td\kappa = 
{i^{-\q\delta +1}\over {2(q-1)}}c\ ,
\ee
where $\nu$ is a constant. Now, using these relations in 
the explicit forms of $A, B, \wtd A$ and $\wtd B$, it is 
easy to relate the other parameters that appear in the $p$-brane and 
cosmological solutions:
\bea 
\tilde \beta &=& {i^{-\delta \q +1}\over (q-1)}\beta, \qquad 
\tilde \mu = {i^{-\delta \q +1}\over (q-1)}\mu\ , \nn\\
\tilde \lambda &=&
i^{-{a^2(D-2)\delta\over{2(q-1)}} +1}\,e^{-{a^2(D-2)\tilde\mu\nu
\over{2(q-1)}}}\lambda, \qquad \alpha = \gamma -i{\pi\over 2}\ . 
\label{map}
\eea

         The above demonstration illustrates that for the case $k=1$
and $\td k=0$, the type 1 cosmological solutions are identical to
those of the type 1 non-extremal p-brane solitons obtained in
\cite{lpx1}, up to general coordinate transformations that may include
Wick rotations.  In fact once the equivalence of the type 1 $p$-brane
ansatz (\ref{bpbrane}) and the cosmological ansatz (\ref{abcmetric})
for $k=1$, $\td k=0$ metrics is established the equivalence of the
solutions is guaranteed, since in each case the most general solutions
compatible with the ansatz were obtained.
Although the non-extremal $p$-branes are not BPS saturated, and thus
can suffer modifications at the quantum level, they do, of course,
have extremal limits in which they become BPS saturated.  This occurs
when the parameter $\td\kappa$ goes to zero.  It is of interest to
enquire what happens to the corresponding cosmological solutions in
this limit. From the constraint given under (\ref{bpsol}), we see that
setting $\td\kappa=0$ in the $p$-brane solutions implies that the
constants $\tilde \beta$ and $\tilde \mu$ must go to zero, and thus
from (\ref{map}) we must have $\beta$ going to zero also.  From
(\ref{psol}), this would imply that $\Phi$ is either singular, or
complex.  Thus there is no limit possible in which the cosmological
solutions can become supersymmetric.  (The fact that they are
non-supersymmetric was observed in \cite{low1}.) It is interesting, 
however, that the
near-extremal regime of non-extremal $p$-branes does map into real
cosmological solutions.  Thus it may be that some of the desirable
properties associated with near-extremality for $p$-branes may have
their analogues in cosmological string solutions.

      So far we have seen that the single-scalar purely electric or
purely magnetic type 1 cosmological solutions can be one-to-one mapped
to the associated type 1 non-extremal $p$-branes.  The
generalisation of this mapping to multi-scalar or dyonic solutions is
straightforward.  The corresponding non-extremal $p$-brane and
cosmological solutions can be found in \cite{lpx1} and \cite{lmpx}
respectively. 

      As we discussed in section 2, cosmological solutions also exist
for other values of $k$ and $\td k$.  As shown above when $\td k =0$,
the corresponding flat space with metric $ds_\q^2$ can be
reinterpreted as the world volume of the $p$-brane.  It seems that
such an interpretation breaks down when $\td k\ne 0$, and such
cosmological solutions would map into solutions that do not have an
interpretation as $p$-branes.  On the other hand, there is no such
restriction on $k$, which defines the structure of the transverse
space of the corresponding $p$-brane.  When $k=1$, we have seen that
the cosmological solutions can be mapped to isotropic $p$-branes.  By
the same token, we can obtain new non-extremal $p$-brane solutions by
mapping the cosmological solutions with other values of $k$. In the
case of $k=0$, the corresponding $p$-brane solutions allow extremal
limits, and were obtained in \cite{dlps}.  These supersymmetric
solutions describe configurations of $p$-branes uniformly distributed
over a $q$-dimensional hyperplane.  They become domain walls after
compactifying these $q$ coordinates \cite{lpdomain}.

       So far, we related type 1 cosmological solutions and the type 1
non-extremal $p$-branes.  As was shown in \cite{hs,dlp1}, there is
another universal way to blacken extremal $p$-branes, namely the type
2 $p$-branes given by (\ref{stand}) and (\ref{standsol}).  It was
observed in \cite{lw} that in the region where $e^{2f}$ is negative,
the signature of the $t$ and $r$ coordinates reverses, leading to a
natural interpretation of the interior $p$-brane solution, inside the
horizon at $e^{2f}=0$, as a cosmological solution.  In this case there
is no need to perform any Wick rotation, since the sign reversal of
the function $e^{2f}$ automatically gives the original $r$ coordinate
a timelike interpretation, while the original time coordinate becomes
spatial.  For the case $p=0$, as we have already remarked, the
standard and the non-standard black $p$-branes coincide, and the
description of the mapping to the cosmological solutions is identical.
Let us therefore consider the remaining cases, when $p>0$.  It seems
to be natural to consider the universe as beginning at $r=0$, since
this is a point where the curvature is singular.  The universe will
then evolve to the horizon at $r=\kappa^{1/\td d}$.  At $r=0$ the
scale factors for $dx_i dx_i$ and for $d\Omega^2$ are zero, expanding
to a finite size at the horizon.  In this interval, the comoving time
$\tau=\im \int^r e^{B-f} dr$ covers a finite range.  The scale factor
for $dt^2$ at the horizon is zero, whilst it is proportional to
$r^{\td d(2a^2/d-1)}$ at small $r$.  Thus to give a satisfactory
cosmological model we must have $a^2\ge d/2$, since there should not
be spatial directions that grow to infinite size at a finite comoving
time.

      To summarise, we have seen that there are two types of
non-extremal $p$-branes solutions: the non-standard (type 1) with the
metric (\ref{bpbrane}), and the standard (type 2) with the metric
(\ref{stand}).  Correspondingly there are type 1 and type 2
cosmological solutions.  We observed that the type 1 black $p$-branes
are identical to the type 1 cosmological solutions that were obtained
in section 2, with $k=1$ and $\td k=0$. In general, the mapping
between these $p$-brane metrics and cosmological metrics requires the
Wick rotation of one or more coordinates.  It is also straightforward
to observe that the interior of the type 2 black $p$-branes can be
interpreted as type 2 cosmological models.  In this case, no Wick
rotation of coordinates is necessary.  It is of interest to study the
inter-relationship of these solutions

         Since the time coordinate $t$ of a type 2 black $p$-brane
becomes spacelike inside the horizon, while the original $r$
coordinate becomes timelike, one can choose to perform a Kaluza-Klein
reduction of the interior metric in which $t$ is compactified.  This
gives rise to a lower-dimensional cosmological solution with the
general form of the type 1 models.  There is however a slight
complication in general, since the lower-dimensional solution will
acquire an additional scalar degree of freedom.  However, if we start
with a higher-dimensional type 2 solution with no dilaton excitation,
its compactification gives precisely a type 1 cosmological
solution. Since this solution can be mapped to a type 1 black
$p$-brane by Wick rotation, it implies that under appropriate
circumstances type 1 black $p$-branes can also obtained from type 2
black $p$-branes by compactifying the time coordinate, made spacelike
by virtue of the Wick rotation.  We shall now illustrate this, taking
the type 2 black membrane in $D=11$ as an example.  Its metric is
given by
\be
ds_{11}^2 = H^{-\ft23}\, (-e^{2f}\, dt^2 + dx^i dx^i) +
            H^{\ft13}\, (e^{-2f}\, dr^2 + r^2 \, d\Omega_7^2)\ ,
\ee
where $e^{2f} = 1 - \kappa r^{-6}$ and $H = 1 + \kappa r^{-6}\,
\sinh^2\mu$. If we simply compactify the solution on one of the
world-volume spatial coordinates $x^i$, it reduces to a type 2 black
string in $D=10$.  However we can instead perform Wick rotations to
make the coordinate $t$ spacelike, and the metric $dx^idx^i$ to be
Minkowskian, and then compactify the Wick rotated $t$ coordinate to
get a different dimensionally-reduced 10-dimensional metric
$ds_{10}^2$, given by $ds_{11}^2= e^{\varphi/6}\, ds_{10}^2 +
e^{-4\varphi/3}\, d(\im t)^2$.  Thus we find
\bea
ds_{10}^2 &=& H^{-\ft34}\, e^{\ft14f}\, dx^i dx^i +
H^{\ft14}\, e^{\ft14f}\, (e^{-2f}\, dr^2 + r^2 \, d\Omega_7^2)\ ,\nn\\
e^{2\varphi} &=& H\, e^{-3f}\ . 
\label{10met}
\eea   
This clearly has the structure of a type 1 black string solution, in
that the metric coefficients for the time direction (one of the two
$x^i$ coordinates) and the spatial world-sheet direction are
identical.  Indeed, it is not difficult to show that (\ref{10met}) is
precisely of the form of the type 1 black metrics (\ref{bpbrane}),
with
\be
\td\kappa = 4\kappa \ ,\qquad \td\mu= -2\kappa\ , \qquad r=\td r
\Big(1+\fft{\kappa}{\td r^6}\Big)^{\ft13}\ .
\ee
Thus we see that the dimensional reduction of the Wick-rotated time
coordinate of a dilaton-free type 2 black $p$-brane gives a special
case of a type 1 black $(p-1)$-brane in one lower dimension. 

    A special case of the type 2 $\longrightarrow$ type 1 reductions
discussed above is when one begins with a black hole in the higher
dimension.  In order to make an exact correspondence with the type 1
solutions, we should again consider examples where there is no dilaton
involved in the higher-dimensional solution.  A simple example is
provided by considering black-hole solutions in the $N=1$ string in
six dimensions, where the charge is carried by a Yang-Mills field.
Including the effects of loop corrections, it was shown in \cite{sag}
that the relevant part of the effective Lagrangian is given by
\be
{\cal L}= e R -\ft12 e\, (\del\phi)^2 -\ft14 (v e^{-\phi/\sqrt2} +
\td v e^{\phi/\sqrt2})\, F^2 \ ,
\ee
where $v$ and $\td v$ are constants.  This admits a black-hole
solution \cite{lmpr} where $\phi$ is a constant, given by 
$e^{\phi\sqrt2}=v/\td
v$, and the metric is given by
\be
ds_6^2= - H^{-2}\, e^{2f}\, dt^2 + H^{\ft23}\,(e^{-2f}\, dr^2 + r^2\,
d\Omega_4^2)\ ,
\ee
where $H=1+ \kappa r^{-3}\, \sinh^2\mu$ and $e^{2f}= 1 -\kappa
r^{-3}$.  Performing a dimensional reduction of the $t$ coordinate
(having either first Wick-rotated it, or else by considering the
interior solution where $e^{2f}$ is negative), we obtain the
5-dimensional solution
\bea
ds_5^2&=& e^{\ft23 f} (e^{-2f}\, dr^2 + r^2\, d\Omega_4^2)\ ,\nn\\
e^{-\sqrt{\ft32}\varphi} &=& H^{-2}\, e^{2f}\ .
\eea
In the exterior region $e^{2f}>0$ this describes a non-extremal
instanton with a Euclidean signature; in the interior region it can be
viewed as a cosmological solution, since the coordinate $r$ becomes
timelike.  Note that the metric is independent of the electric charge,
which appears only in the harmonic function $H$.  In fact this is a
general phenomenon that will occur whenever a black-hole solution is
compactified in the time direction.  Thus starting from the black-hole
metric
\be
ds_{\sst D}^2 = -H^{-\ft{4(D-3)}{\Delta(D-2)}}\, e^{2f}\,dt^2 +
H^{{4}{\Delta(D-2)}} \, (e^{-2f}\, dr^2 + r^2\, d\Omega_{\sst D-2}^2)\ ,
\ee
we reduce it to $(D-1)$ dimensions by applying the standard
Kaluza-Klein reduction on the time coordinate, $ds_{\sst D}^2 =
e^{2\a\varphi}\, ds_{\sst D-1}^2 - e^{-2(D-3)\a\varphi}\, dt^2$,
implying that
\bea
ds_{\sst D-1}^2 &=& e^{\ft{2f}{D-3}}\, (e^{-2f}\, dr^2 + r^2\,
d\Omega_{\sst D-2}^2)\ ,\nn\\
e^{\a\varphi} &=& H^{\ft{2}{\Delta(D-2)}}\, e^{-\, \ft{f}{D-3}}\ ,
\eea
where $\a^{-2}= 2(D-2)(D-3)$.

\vfill\eject

\end{document}